\documentclass{ws-procs9x6}

\begin{document}

\title{The Pierre Auger Observatory: Science Prospects and Performance at First Light}

\author{Luis A. Anchordoqui}
\author{[\lowercase{for the} AUGER C\lowercase{ollaboration}]}

\address{Department of Physics \\
Northeastern University, \\ 
Boston MA 02115\\ 
E-mail: l.anchordoqui@neu.edu}


\maketitle

\abstracts{The Pierre Auger Observatory is a major international effort aiming at high-statistics study of 
highest energy cosmic rays. A general description of the experimental set-up and overall performance of 
the detector at first light are presented.}

The Pierre Auger Observatory (PAO) is designed to study cosmic rays with energies above about 
$10^{10}$~GeV with the aim of uncovering their origins and nature.\cite{Abraham:2004dt}
Such events are too rare to be directly detected, but the direction,
energy, and to some extent the chemical composition of the primary particle
can be inferred  from the cascade of secondary particles induced 
when the primary impinges on the upper atmosphere.\cite{Anchordoqui:2002hs} 
These cascades, or air 
showers, have been studied in the past by measuring the nitrogen fluorescence they produce in the
atmosphere,\cite{Baltrusaitis:1985mx} or by directly sampling shower particles at ground 
level.\cite{Chiba:1991nf}  The PAO
is a ``hybrid'' detector, exploiting both of these methods by employing an array of water
\v{C}erenkov detectors overlooked by fluorescence telescopes; on clear, dark nights
air showers are simultaneously observed by both types of detectors,  
facilitating powerful reconstruction methods and control of the systematic
errors which have plagued cosmic ray experiments to date.  
Additionally, since the center--of--mass energy in the collision of the primary with the atmosphere is 
above about 100 TeV ({\it i.e.,} exceeding the contemporary and forthcoming collider reach by 
2 orders of magnitude), PAO will also be capable of probing new physics beyond the electroweak 
scale.\cite{Anchordoqui:2003zk}

The Observatory will be covering two sites, in the Southern (Pampa Amarilla) and 
Northern hemispheres. Each site consist of 1600 stations spaced 1.5 km apart from each 
other, with 4 fluorescence eyes placed on the boundaries of the surface array. The 
energy threshold is defined by the 1.5~km spacing of the detector stations: a $10^{10}$~GeV vertical 
shower will hit on average 6 stations which is enough to fully reconstruct the extensive air shower.
The installation of the Southern site is now well underway, with detectors
looking at the yet poorly covered part of the sky in which the direction of the center of the Milky 
Way is visible.

\begin{figure}[ht]
\centerline{\epsfxsize=4.1in\epsfbox{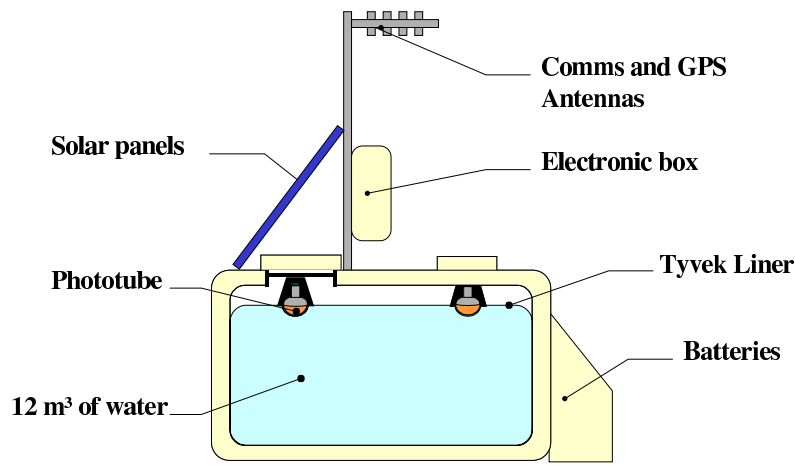}}   
\caption{Schematic view of a water \v{C}erenkov detector. \label{tank}}
\end{figure}

Each ground-based detector is a cylindrical opaque tank of 3.6~m diameter and 1.55~m high, 
where particles produce light by \v{C}erenkov 
radiation, see Fig.~\ref{tank}.
The filtered water is contained in a bag which diffusely
reflects the light collected by three photomultipliers (PMT's) installed on the
top. The large diameter PMT's ($\approx$ 20~cm) 
are mounted facing down and look at the water through sealed polyethylene 
windows that are integral part of the internal liner. 

Due to the size of the array, the detectors have to be able
to function independently. The stations operate on battery-backed solar power and 
time synchronisation relies on a comercial Global Positioning System (GPS) receiver.\cite{Pryke:1995uv}
A specially designed wireless LAN radio system is used to provide communication between 
the surface detectors and the central station.\cite{Clark}
Each tank forms an autonomous unit, 
recording signals from the ambient cosmic ray flux, independent of the signals 
registered by any other tank in the surface detector array. A combination of signals clustered in 
space and time is used to identify a shower.

\begin{figure}[ht]
\centerline{\epsfxsize=3.1in\epsfbox{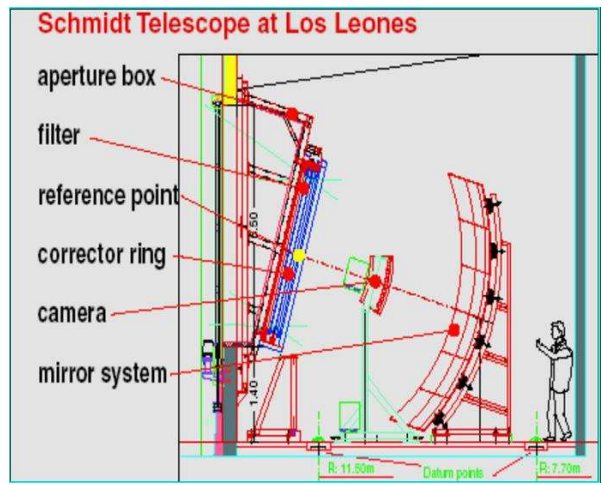}}   
\caption{Schematic view of a fluorescence telescope at Los Leones. From left to right can be seen the 
aperture system, the photomultiplier camera and the spherical mirror.\label{eye}}
\end{figure}


A single fluorescence detector unit (or eye) comprises 6 telescopes, each located in 
independent bays, overlooking separate volumes of air. A 
schematic cross-sectional view of one fluorescence telescope is shown in Fig.~\ref{eye}. A 
circular diaphragm, positioned at the centre of curvature of the spherical mirror, 
defines the aperture of the Schmidt optical system.\cite{Cordero} Ultra-violet transmitting filters are 
installed in the entrance aperture. Just inside the ultra-violet filter is a ring of (Schmidt) 
corrector elements. Light is focused by a large 3.5~m $\times$ 3.5~m spherical mirror onto a 
440 PMT camera, which accommodates the $30^\circ$ azimuth $\times$ $28.6^\circ$ elevation field of view.  
Each camera pixel has a field of view of approximately $1.5^\circ$.

In the hybrid mode ($\sim 10\%$ of the time) the detector is expected to have energy resolution of $13\%$ 
$(1\sigma)$ at $10^9$~GeV improving to 5.5\% at $10^{11}$~GeV, and an angular resolution of about 
$0.5^\circ.$ For the ground array alone these numbers become 10\% and $1^\circ,$ again for primary 
energy $> 10^{11}$~GeV. Estimating event rates is a risky business because above $10^{11}$~GeV the 
fluxes are essentially not known.\cite{Takeda:2002at}
However,  for zenith angle less than  60$^\circ$ the total aperture of the Southern surface array is 
$\sim 7350$~km$^2$ sr, and thus extrapolation from AGASA measurements\cite{Takeda:2002at} implies 
that PAO should 
detect of the order of 2500 events above $10^{10}$~GeV and of a 50 to 100 events above $10^{11}$~GeV 
every year. Moreover,
if events
above 60 degrees can be analyzed effectively, the aperture will increase by
about 50\% .

PAO also provides a promising way of detecting ultra-high energy neutrinos by looking for 
deeply--developing, large zenith angle ($>60^\circ$) or horizontal air showers.\cite{Capelle:1998zz} At 
these large angles, hadronic 
showers have traversed the equivalent of several times the depth of the vertical atmosphere and their 
electromagnetic component has extinguished far away from the detector. Only very high energy core--produced 
muons survive past 2 equivalent vertical atmospheres. Therefore, the shape of a hadronic (background in 
this case) shower front is very flat and very prompt in time. In contrast, a neutrino shower appears pretty 
much as a ``normal'' shower. It is therefore straightforward to distinguish neutrino induced events from 
background hadronic showers. Moreover, if full flavor mixing is confirmed, tau neutrinos could be as 
abundant as other species and so very 
low $\nu_\tau$ fluxes could be detected very efficiently by PAO's detectors by looking at the 
interaction in the Earth crust of quasi horizontal $\nu_\tau$ inducing a horizontal cascade at the 
detector.\cite{Bertou:2001vm}

\begin{figure}[ht]
\centerline{\epsfxsize=4.1in\epsfbox{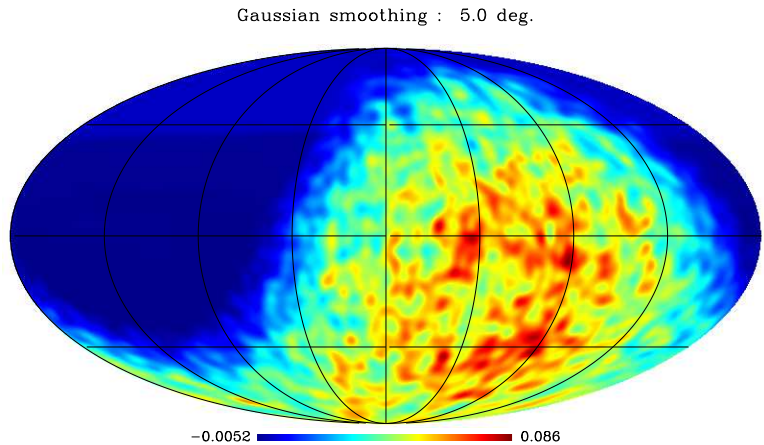}}   
\caption{All event (from January to July 2004) skymap in Galactic coordinates. Units $x$ are related to the 
number of events per pixel $n$ according to: $n = 330 \,x + 1.716.$ \label{skymap}}
\end{figure}

The first PAO site is now operational in Malarg\"ue, Argentina, and is in the process
of growing to its final size of 3000~km$^2$.
At the time of writing, 12 telescopes and about 400 water tanks were operational. 
The first analyses of data from the PAO are currently underway. Figure~\ref{skymap} shows the 
arrival directions of all events recorded from January to July 2004. The pixels have a size of 1.8 
degrees and the map was smoothed with a Gaussian beam of 5 degrees.\cite{Armengaud} On 21 May 2004, 
one of the larger events recorded by the surface array triggered 34 stations. A preliminary estimate 
yields an energy~$\sim~10^{11}$~GeV and a zenith angle of about $60^\circ.$ First physics results 
will be made public in the Summer of 2005 at the 29th International Cosmic Ray Conference.

\section*{Acknowledgments}

I would like to thank all my collaborators in the Pierre Auger
Observatory. Special thanks goes to Tere Dova, Jean-Christophe Hamilton, 
Antoine Letessier-Selvon, Tom McCauley, Tom Paul, Steve Reucroft, and 
John Swain for assistance in the preparation of this talk. This work has 
been partially supported by the US National Science Foundation (NSF), under grant No.\
PHY--0140407.


\noindent  The references to internal reports written by the AUGER Collaboration 
(GAP notes) are accesible via {\tt www.auger.org/admin/GAP$_{-}$NOTES}.

\end{document}